# Software Performance Analysis

Michel R. Dagenais (michel.dagenais@polymtl.ca)
Dept. of Computer Engineering, Ecole Polytechnique, Montreal, Canada

Karim Yaghmour (karim@opersys.com)
Opersys, Montreal, Canada

Charles Levert (Charles.Levert@ericsson.ca)
Makan Pourzandi (Makan.Pourzandi@ericsson.ca)
Ericsson Research, Montreal, Canada

## 1. Abstract

The key to speeding up applications is often understanding where the elapsed time is spent, and why. This document reviews in depth the full array of performance analysis tools and techniques available on Linux for this task, from the traditional tools like gcov and gprof, to the more advanced tools still under development like oprofile and the Linux Trace Toolkit. The focus is more on the underlying data collection and processing algorithms, and their overhead and precision, than on the cosmetic details of the graphical user interface frontends.

## 2. Introduction

Code coverage tools (e.g., Gcov [1]) compute the exact number of times each statement is executed, and thus can easily pinpoint frequently executed sections. However, code coverage tools carry a non negligible execution overhead, and more importantly do not measure the relative execution cost of different statements. Time domain sampling tools (e.g., Gprof [2]) may therefore more accurately report large contributors to the execution time by sampling at regular time intervals the execution address. Moreover, the overhead cost may be traded off for accuracy by varying the time interval.

Modern applications and systems present new challenges as their performance may be strongly affected by cache hit rates, disk accesses, inter-process communications, synchronization waits, and operating system overhead. This is where new tools based on hardware performance counters sampling (e.g., Oprofile [3]), or detailed analysis of low overhead operating system traces (e.g., Linux Trace Toolkit [4]) can bring more information. They may for instance indicate code sections with exceedingly high level of cache misses or branches mispredictions, or estimate the execution time of the X window server [5] on behalf of each of its clients.

Section 3 discusses the algorithms and techniques used to collect, process and store performance data. The following section describes a number of specific performance analysis tools, from commonly used tools to more recent ones, with a longer section on possible extensions and customisations to the Linux Trace Toolkit. Section 5 is the conclusion.

## 3. Performance Data Collection, Processing and Storage

In a simple model, a piece of performance data is obtained, and queued or stored for later display. In practice, the generated volume of data may be such that some processing may be required at data collection time to synthesize more compact information. For example, in a large scientific computer cluster of 1024 computers, each time a program function is called, the entry and exit time may be collected. Sending these values directly to a central server would rapidly saturate the network and carry an unbearable overhead. Instead, the entry and exit time may be subtracted and accumulated locally in each function, in order to produce a summary of the processing time spent in each program function. Individual programs, upon exit, would send the execution time histogram to a central server for offline preparation of performance reports.

This section examines the key aspects of data collection and processing: obtaining the data points, accumulating values in temporary storage, and sending the values to storage for offline processing and display.

## *3.1. Data Collection*

Several activities including debugging and program understanding require data collection, just like performance analysis, and use the same techniques. The familiar debugging process is used to illustrate common data collection methods.

There is a compromise between the execution overhead and the amount and precision of the data collected. Sometimes the overhead is unacceptable because the resulting program execution time becomes too large, or the program performance or the real time program behavior is skewed by the data collection.

Storing or summing a value in a variable is minimal overhead. Calling another function for each function entry is a large overhead for small functions. Writing performance data to a file upon entering a basic block (i.e., linear code section, without any jump, and thus always executed together and the same number of times) would significantly change the execution performance of a program.

## 3.1.1. Manual Program Instrumentation

Adding a print statement is often the easiest and most flexible way to obtain the desired information. All the data structures and functions accessible to the program may be used to generate the data printed. The Linux kernel code, where novice programmers are uncommon, contains no less than 80000 printk statements. Several common tools (netstat, vmstat, mtrace) use data gathering statements inserted in the kernel, or the C librabry.

## 3.1.2. Automatic Program Instrumentation

Compilers and preprocessing tools may be used to systematically instrument a program by adding data gathering statements to the code they would normally generate. Typical locations for inserting statements include the functions entry and exit, and the entry of linear code sections (basic blocks).

For example, the GNU Compiler Collection, GCC [1], offers several code instrumentation options to insert a call to function *mcount* upon entering each function (option -pg used for gprof), and to insert calls to hooks upon entering linear code sections and jumping to other code sections (options -ftest-coverage and -fprofile-arcs which may be used for gcov).

## 3.1.3. Binary Instrumentation

In some situations, the source code is not available for source level instrumentation and recompilation, or it is simply inconvenient. There are tools to modify binary executable files in order to insert data gathering statements (e.g., EEL [6]).

Ideally, the existing binary instructions should be shifted a few bytes higher in the address space each time instrumentation instructions need to be inserted. Doing so, however, changes all the addresses in the program and all jumps and calls must be found and adjusted accordingly. Unless the program section containing instructions does not contain any data (e.g., jump tables) and all instructions have the same size, this rapidly gets very tricky.

Another easier technique is code patching. The instruction(s) located where instrumentation code should be inserted is replaced by a jump to a new code section appended to the program, and containing the instrumentation code and the replaced instruction. Only the one or two instructions overwritten by the jump need to be relocated (address adjusted) per instrumentation point.

A simpler form of code patching is used in most debuggers to insert breakpoints. The instruction where instrumentation code should be inserted is saved away and replaced by a soft interrupt instruction. Unlike jump instructions which are among the longuest, and may overwrite a few shorter instructions, interrupt instructions usually are the smallest and occupy a single byte on i386 processors. Once the interrupt instruction is reached, the interrupt handler executes the instrumentation code, puts back the saved away instruction and executes it in single stepping mode. This way, another interruption is generated after its execution, at which point the interrupt instruction is placed back for the next time the program reaches this point.

With this technique, one soft interrupt and one single stepping interrupt is required for each instrumentation point. The address of the instruction causing the interrupt is used to determine the instrumentation point reached and the associated code to execute. This least invasive technique (no instructions to relocate) also carries the largest overhead. Interrupts are processed in microseconds at best while jumps are executed in nanoseconds, as shown below with GDB [7].

```
# Run a program under gdb
[gzip-1.2.4a]$ cat gdb.run
file gzip.plain
run </tmp/evlogout >/dev/null
quit

[gzip-1.2.4a]$ time gdb -command=gdb.run >gdb.out
23.30user 1.53system 0:30.75elapsed 80%CPU (0avgtext+0avgdata 0maxresident)k
0inputs+0outputs (1213major+875minor)pagefaults 0swaps

# Run a program under GDB with a breakpoint. The breakpoint is executed
# 6933680 times, each time printing a "." before continuing.
[gzip-1.2.4a]$ cat gdb.commands
file gzip.plain
break deflate.c:679
commands 1
silent
print "."
continue
end
run </tmp/evlogout >/dev/null
quit

#Run GDB with the command file
[gzip-1.2.4a]$ time gdb -command=gdb.commands >gdb.out
3905.78user 1535.75system 2:02:48elapsed 73%CPU
```

### 3.1.4. Sampling

Sampling is often used to predict at low cost the outcome of an election by asking several hundred people their voting preferences instead of calling each of the millions eligible voters. Similarly, instead of recording the execution time for each function, the program may be interrupted at regular intervals (e.g., each 10 miliseconds) to sample the address of the currently executed instruction. The chances of the address being within a specific function is proportional to the proportion of time spent in that function (under randomness assumptions).

Simple sampling based programs like gprof use imprecise virtual timers which are checked each time the operating system scheduler is run. In programs with strong interactions with the scheduler (e.g., frequently polling for I/O or calling *yield*), the results may be severely skewed. Newer tools like oprofile take advantage of hardware performance counters which may count different events (cycles, cache misses, branches taken/not taken...), and generate an interrupt after a certain count. This allows varying easily the sampling interval, trading off overhead for accuracy, and measuring several different parameters apart from execution time. The low achievable overhead enables measuring the cache misses histogram for a program, which would be impractical using code instrumentation.

## 3.2. Performance Data Processing

Once a data element is obtained, it must find its way to the results file. Tracing systems usually add a few fields (record number, time, address...) to the data element and append it to a buffer which is written to disk once full. This way the complete information is retained but the disk may fill at an alarming rate. Tracing the system calls for a Web browser may generate several megabytes per minute.

The alternative to simple buffering is to process and aggregate the data elements as they are generated. Typically a counter variable will be associated with each function or basic block, and the events occuring there will be counted as they happen. At the program end, or at regular intevals, the content of these counters is written to disk. In some cases, the number of possible locations is much larger than the number actually used and hash tables are used. This is the case when counting the calls to a function by caller. Very few functions among all the functions in a program actually call a specific function; using hash tables is thus more efficient than allocating a counter for each possible caller-callee combination.

## 3.3. Performance Data Transfer and Storage

In low volume applications, events are generated every few seconds or more, and may be written as they happen to a file descriptor. This file descriptor may be connected to a file, a socket or even a graphical viewer which updates its display in real time. For larger volumes of events, events should be appended in memory to a large buffer, written to disk in very large chunks. When the instrumented program is different from the logging process writing to disk, maximum performance may be obtained through the use of shared memory between the instrumented program and the logging daemon. The Linux Trace Toolkit uses shared memory between the instrumented kernel and the tracing daemon to achieve maximum performance.

The performance parameters being measured (disk throughput, execution time, cache hits) will strongly influence which overhead should be minimized. If disk access times are to be measured, the impact of the performance data gathering on disks should be reduced, for instance by writing to a separate disk, on a separate disk controller, and even across the network on a separate computer. If only one disk is available, it is also possible to reduce the disk requirements at the expense of execution time using data compression. Most systems perform data compression offline for archival purposes but not during the live tracing.

# 4. Review of Performance Analysis Tools

The tools reviewed in this section are all freely redistributable and cover the full spectrum of widely used techniques for software performance analysis. The simpler and more widely deployed tools are presented first. Throughout this section, the GNU compression utility gzip [8] will be used as a sample test case, compressing a 64MB log file to 1.6MB. It is used to compare the overhead of different performance analysis tools.

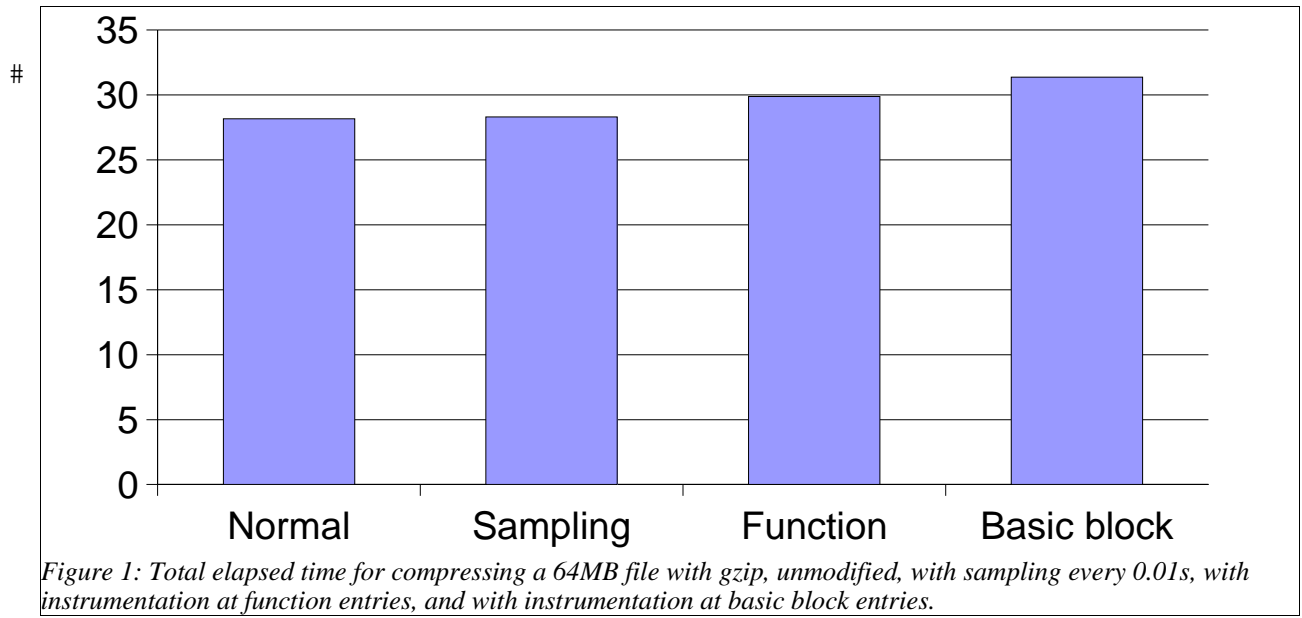

*Figure 1: Total elapsed time for compressing a 64MB file with gzip, unmodified, with sampling every 0.01s, with instrumentation at function entries, and with instrumentation at basic block entries.*

```
Run the normal program
[gzip-1.2.4a]$ time ./gzip.plain </tmp/evlogout >/dev/null
24.09user 0.72system 0:28.16elapsed 88%CPU
```

## *4.1. Code instrumentation and results processing with GCC/GCOV*

Gcov is part of the standard GNU C Compiler (GCC) distribution [1]. When compiling a program foo.c with the -ftest-coverage and -fprofile-arcs options, a list of the basic blocks and branches are generated in files foo.c.bb and foo.c.bbg, and the code is instrumented to count the execution frequencies of each basic block and branch and to write out the results upon exit. The program may then be executed and sums its results in file foo.c.da, which can later be analyzed with gcov to print the source code along with execution frequencies.

This tool has a very low overhead, a single *add* instruction for each basic block and each branch execution. The basic block execution frequencies are used to detect frequently used code sections, where optimisation coding efforts may be worthwhile. Execution frequencies for basic blocks and branches may be used for test coverage analysis. Any section with null execution count has not been covered by the tests used when executing the program. Finally, branches counts may be used to compute branch probabilities, which may be used in subsequent compilations to generate better code for the more frequent case.

Another GCC option, -ax produces similar information with more control over the generated data but without a post-processing tool like Gcov. It can produce functions, basic blocks and branches execution counts, or a full trace of function and basic block entries. Furthermore, it is possible to specify functions to exclude/include from this profiling/tracing.

As an example, the GNU compression program gzip was recompiled with the -ftest-coverage and -fprofile-arcs options and run on a 64MB log file. The instrumented gzip requires 5.4% more elapsed time than the normal gzip. At the end of the execution, a file with the .da extension is created for each source code file (e.g., deflate.da for deflate.c). These files may then be used by gcov to produce annotated source code files (e.g., deflate.c.gcov for deflate.c).

```
# Run the program compiled and linked with options
# -ftest-coverage and -fprofile-arcs
[gzip-1.2.4a]$ time ./gzip.gcov </tmp/evlogout >/dev/null
27.04user 0.63system 0:31.36elapsed 88%CPU

# Files .bb and .bbg are produced at compilation time, files .da
# at the program exit.
[gzip-1.2.4a]$ ls -als deflate.*

  4 -rw-rw-r--    1 dagenais dagenais     1100 avr  9 15:34 deflate.bb
  4 -rw-rw-r--    1 dagenais dagenais     2048 avr  9 15:34 deflate.bbg
 32 -rw-r--r--    1 dagenais dagenais    29179 aoû 13  1993 deflate.c
  4 -rw-rw-r--    1 dagenais dagenais      648 avr 10 14:10 deflate.da
 20 -rw-rw-r--    1 dagenais dagenais    18856 avr 10 12:01 deflate.o

# Perform the analysis on one of the source files.
# Gcov reads the .bb .bbg .da and .c files and produces .c.gcov
[gzip-1.2.4a]$ gcov -b -f deflate.c
 80.77% of 26 source lines executed in function lm_init
 60.00% of 15 branches executed in function lm_init
 40.00% of 15 branches taken at least once in function lm_init
No calls in function lm_init
...
Creating deflate.c.gcov.

[gzip-1.2.4a]$ less deflate.c.gcov
...
            5          while (lookahead != 0) {
branch 0 taken = 0%
      6933680            INSERT_STRING(strstart, hash_head);
      6933680            prev_length = match_length, prev_match = match_start;
      6933680            match_length = MIN_MATCH-1;
      6933680            if (hash_head != NIL && prev_length < max_lazy_match &&
branch 0 taken = 8%
branch 1 taken = 47%
branch 2 taken = 1%
                          strstart - hash_head <= MAX_DIST) {
      3367555            match_length = longest_match (hash_head);
      3367555            if (match_length > lookahead) match_length =
lookahead;
branch 0 taken = 0%
branch 1 taken = 100%
...
```

## *4.2. Time Based Sampling with GPROF*

Gprof is part of the GNU binutils package [2]. When a program foo.c is compiled with the -pg option, a call to function *mcount* is added to each function entry. Moreover, the link step uses versions of the standard libraries compiled with -pg when available, and adds a module to initialize the sampling at the beginning and to write out the results at the end. Optionally, compilation flag -g may be specified to obtain source code lines addresses (usually used by debuggers), and -a to simultaneously compute basic block execution frequencies.

When the program executes, an interrupt handler is called every clock tick (10 miliseconds) which takes the address of the interrupted instruction and adds 1 to the corresponding bucket in a large array. Typically this array is as large as the process instructions area and uses a 4 bytes integer counter for each 4 bytes section in the program. Each time a function is called, the internal *mcount* function gets called, extracts the addresses of the caller and the callee, and sums the number of calls for this caller-callee pair in a large hash table. When the program exits, the samples counts in the large array, and the caller-callee counts from the hash table, are printed in a results file named gmon.out. When the -a compilation flag is specified, basic blocks execution counts are also computed and are appended to the results file.

The Gprof program may then be used to read results files, and the program executable and source code files, to produce different reports. The flat profile report prints for each function (and even each source code line) the number of associated samples and the corresponding execution time probably spent in that location. The functions may be printed in decreasing order of execution time, thus listing on top the most time consuming functions, and best candidates for optimisation. The time associated with internal function *mcount* can be used to substract, and thus compensate, the overhead associated with the profiling.

The annotated source code report is similar to the output produced by Gcov, listing the execution frequencies for each basic block. Finally, The call graph report is a superset of the flat profile and adds information about callers and called functions. For each function, it lists the total time spent in itself (number of samples times sampling period), and the calling functions. For each calling function is listed the number of calls and the proportion of the total; this information is obtained precisely through the calls to *mcount* where the caller is recorded in the call counts. The proportion of calls from a caller is used to estimate the time spent in the current function by caller. The underlying assumption is that calls to a function take the same time on average irrespective from the caller. This is clearly not the case as, for instance, a function draw_icon may call scale_bitmaps for very small images while function scale_background may call scale_bitmaps for images 500 times larger.

The execution time spent for each caller is propagated to the callers resursively, starting from the leaf functions. This adds another list for each function in the call graph report, the time spent in each function (and its children) called. The total time spent in the function itself and all its called children is also part of the call graph report. For recursive calls, a separate entry is created for the cycle which lists callers and callees from ouside the cycle; then, entries for members of the cycle list callers and callees from within the cycle.

Gprof and similar tools offer an excellent compromise of low overhead and detailed usually reliable information. It may, however, be inadequate or even misleading in more involved cases where:

- the time spent in a function is very different depending on the caller. Measuring it adequately would require sampling not only the currently executing instruction address but also the callers address on the stack, possibly for several levels. The call graph report could then show the time spent in a function based on chains of callers, but this would be more difficult to present clearly, and the number of samples associated with each chain is likely to be too small to be reliable;
- the sampling interval is correlated with the program activity. When a program continuously makes blocking system calls, there is a chance that the scheduler often gives control to the program and activates the virtual timer used for sampling when the program is in the same area, which gets overrepresented in the samples;
- the time spent not executing the program, waiting for the disk or for other processes, is of interest. In that case, other sources of information should be used to annotate the reports with information about time spent waiting, to obtain the total elapsed time, instead of just the execution time, in each function;

- more information is required about why time is spent in each section (cache faults, page faults...).

```
# Run the program compiled without -pg and linked with -pg
# to measure the low overhead of sampling alone.
[gzip-1.2.4a]$ time ./gzip.pg-nomcount </tmp/evlogout >/dev/null
24.30user 0.62system 0:28.30elapsed 88%CPU

# Run the program compiled and linked with option -pg
# Overhead is slightly larger but it provides call graph info
[gzip-1.2.4a]$ time ./gzip.pg-static </tmp/evlogout >/dev/null
25.79user 0.51system 0:29.88elapsed 88%CPU

# File gmon.out is produced when the program exits
[gzip-1.2.4a]$ ls -als gmon.out
  188 -rw-rw-r--    1 user user    186366 avr 10 14:07 gmon.out

# Use gprof to interpret gmon.out, using debugging symbols
# information from the executable file
[gzip-1.2.4a]$ gprof gzip.pg-static >gprof.out

# Look at the results, first the flat profile with the percentage of
# execution time for each function in decreasing order.
[gzip-1.2.4a]$ less gprof.out

Each sample counts as 0.01 seconds.
  %   cumulative    self              self    total
 time   seconds   seconds    calls  ms/call  ms/call  name
 27.13     7.53      7.53     1957     3.85     5.95  fill_window
 23.20    13.97      6.44        1  6440.00 19205.66  deflate
 14.84    18.09      4.12     1958     2.10     2.10  updcrc
 12.16    21.46      3.38                             do_scan
  8.38    23.79      2.33                             short_loop
  3.78    24.84      1.05                             __mcount_internal
  3.13    25.71      0.87                             read
...

# Then the call graph. Function updcrc uses 4.12s and calls none.
# It is called from zip and file_read. Function file_read is
# called from lm_init and fill_window. It uses about no time but
# calls updrc for about 4.12s. This 4.12s is almost all attributed
# to caller fill_window (1957/1958 calls).
...
-----------------------------------------------
                0.00    0.00       1/1958        zip [3]
                4.12    0.00    1957/1958        file_read [8]
[7]     15.6   4.12    0.00        1958      updcrc [7]
-----------------------------------------------
                0.00    0.00       1/1958        lm_init [27]
                0.00    4.12    1957/1958        fill_window [6]
[8]     15.6   0.00    4.12        1958      file_read [8]
                4.12    0.00    1957/1958           updcrc [7]
-----------------------------------------------
...
```

## 4.3. Hardware Counted Events Sampling with Oprofile

Oprofile [3] is very similar to Gprof; it even produces compatible results files. The two important differences are that it collects samples based on hardware performance counters interrupts, and that it operates on the complete system (gathering samples continuously for all programs executing on the system).

Hardware performance counters are now available in all powerful modern microprocessors and may be used to count various events of interest including clock cycles, L1 or L2 cache misses, branches, branches taken, and pipeline stalls. The periodic sampling is obtained by requesting an interrupt when the counter reaches a certain count.

Oprofile does not use program instrumentation (i.e., placing calls to *mcount* in each function). Thus, no recompilation is necessary and the overhead is lower, but call graph reports cannot be generated. The only requirement is that debugging information is required to correlate the address of samples with program functions and line numbers. Because of its low overhead, it is not uncommon to keep Oprofile running for extended periods on running server systems.

The whole system profiling feature is particularly useful to see which proportion of the time is spent in each program, thus discovering who is using up all the resources. It is also useful when a single task involving several programs is executed on a system. For example, when analysing the startup time of a sophisticated Web browser, it is possible to see the time spent in functions in the kernel, the browser, the window manager, the X server and various daemons.

In some rare cases, the information produced by hardware performance counters sampling tools may be biased. Indeed, a real-time process using the same period as the sampling tool may have its samples grossly under or over represented. A simple solution is to randomly vary the sampling period around a target average.

```
# Run the oprofile graphical user interface, click to start the daemon
[gzip-1.2.4a]$ oprof_start

# Run the unmodified program (with debugging symbols present)
[gzip-1.2.4a]$ time ./gzip.plain </tmp/evlogout >/dev/null
24.40user 0.87system 0:27.67elapsed 91%CPU

# Stop the profiling daemon and extract statistics
[gzip-1.2.4a]$ oprofpp -l -i gzip.plain >gzip.oprof

# Look at the results
[gzip-1.2.4a]$ less gzip.oprof
Cpu type: PII
Cpu speed was (MHz estimation) : 233.140000
Counted "clocks processor is not halted", sampling interval 200000
vma       samples    %-age         symbol name
08048100 0          0             _start
...
08050504 50         0.059034      long_loop
080504e1 68         0.0802862     limit_ok
080504c1 111        0.131055      longest_match
08050572 160        0.188909      the_end
08050547 171        0.201896      mismatch
0804c550 988        1.16651       ct_tally
0804c930 1100       1.29875       send_bits
0804c6d0 1282       1.51363       compress_block
0805aad0 1465       1.7297        memcpy
0805050d 6905       8.15259       short_loop
08050522 12101      14.2874       do_scan
0804ea10 13821      16.3182       updcrc
0804b4c0 21405      25.2724       deflate
0804b120 25020      29.5406       fill_window
```

## 4.4. System Calls Tracing using Strace

Linux and POSIX systems offer the ptrace system call to allow one process to monitor another process. It is used by debuggers and other tools like strace [9]. When a program is run under strace, it is blocked each time it issues a system call. The control passes to strace before the system call, to extract the value of the arguments, the system call is then performed as usual, and the control passes to strace after the system call to extract the return values. The overhead cost is several microseconds for each system call, comparable to hitting a breakpoint in a debugger. However, the number of system calls is relatively low for CPU bound processes like gzip.

```
# Trace the system calls issued during the execution
[gzip-1.2.4a]$ time strace -s 120 -f -F -o trace ./gzip.plain \
    </dev/evlogout >/dev/null
25.23user 1.57system 0:30.97elapsed 86%CPU

# Examine the system call trace file
[gzip-1.2.4a]$ less trace
2488  execve("./gzip.plain", ["./gzip.plain"], [/* 44 vars */]) = 0
2488  fcntl64(0, 0x1, 0, 0xbffff804)    = 0
2488  fcntl64(0x1, 0x1, 0x1, 0xbffff804) = 0
2488  fcntl64(0x2, 0x1, 0x1, 0xbffff804) = 0
2488  uname({sys="Linux", node="rocamadour", ...}) = 0
2488  geteuid32()                         = 5363
2488  getuid32()                          = 5363
2488  getegid32()                         = 1105
2488  getgid32()                          = 1105
2488  brk(0)                              = 0x810b374
2488  brk(0x810b394)                      = 0x810b394
2488  brk(0x810c000)                      = 0x810c000
2488  rt_sigaction(SIGINT, {SIG_IGN}, {SIG_DFL}, 8) = 0
2488  rt_sigaction(SIGINT, {0x804aaa0, [INT], SA_RESTART|0x4000000}, {SIG_IGN}, 8) = 0
2488  rt_sigaction(SIGTERM, {SIG_IGN}, {SIG_DFL}, 8) = 0
2488  rt_sigaction(SIGTERM, {0x804aaa0, [TERM], SA_RESTART|0x4000000}, {SIG_IGN}, 8) = 0
2488  rt_sigaction(SIGHUP, {SIG_IGN}, {SIG_DFL}, 8) = 0
2488  rt_sigaction(SIGHUP, {0x804aaa0, [HUP], SA_RESTART|0x4000000}, {SIG_IGN}, 8) = 0
2488  ioctl(1, 0x5401, 0xbffff6c0)       = -1 ENOTTY (Inappropriate ioctl for device)
2488  fstat64(0, {st_mode=S_IFREG|0644, st_size=64140109, ...}) = 0
2488  read(0, "recid=1, size=8, format=BINARY, event_type=4, facility=LTT, severity=DEBUG, \nuid=dagenais, gid=dagenais, pid=1953, pgrp="..., 65536) = 65536
...
```

## 4.5. Low Overhead Kernel Tracing with the Linux Trace Toolkit

The data aggregation performed by the tools discussed in previous sections is important to obtain a low overhead. The downside is that some information is lost in the aggregates (number of calls for each function instead of which function was called when and in what order). A full trace of collected data items rapidly fills a disk but does not *a priori* discard information that may not seem useful initially, but could be found so later. Tracing is used pervasively in software projects, from print statements to logging daemons like syslogd and klogd.

The Linux Trace Toolkit [4] is extensible. It comes with 60 predefined tracepoints inserted in the kernel, divided into 8 categories which may be enabled/disabled either at compile or at run time. More tracepoints may easily be added by recompiling the kernel or using the Dynamic Probes code patching system [10]. Its key features are low overhead and post-processing tools.

Low overhead is achieved by carefully selecting the information to log, and minimizing context switching. Since all scheduling changes and setuid calls may be traced, there is no need to record the process or user id with each event. Furthermore, most events being closely spaced in time, 32 bits time differences are stored instead of full 64 bits seconds/microseconds values. All the events are logged by appending their data to a large double buffer. When a buffer is full, a daemon accesses the buffer through shared memory and copies its content to disk. The overhead of tracing a busy system amounts to less than 5% for the elapsed time and up to several megabytes per minute of tracing data to store on disk [11].

The trace analysis is performed offline, to minimize the overhead during data collection, and possibly on a remote host when studying embedded systems. The information contained in the trace (system calls, devices operations, scheduling changes) is sufficient to reconstruct the system behavior (including which file descriptor is associated to which file or memory region) and may therefore be used for several different analyses. The tools included show the events graphically, per process in a timeline, and contain per process summaries (execution time, number of calls per system call, wait time, bytes read/written).

New analysis tools may be written using the provided trace reading library, or by modifying the provided tools. Indeed, while it is easy with LTT to capture all the needed raw data, synthesizing the desired high level information is more challenging. As a test, a modified trace analysis tool was developed to answer the following non obvious question: where is the large starting time of applications such as Galeon, Mozilla and OpenOffice spent?

### 4.5.1. Decomposing the Waiting Time

Traditionally, system statistics decompose the elapsed time into user mode execution time, system mode execution time, and waiting time. Profiling tools are good at decomposing the execution time by function or even source code line. The objective here was decomposing the waiting time into waiting runnable (waiting for the CPU) and waiting non runnable (blocked on I/O, or waiting for a process or a timer). This information may be extracted from a detailed trace:

- Scheduling Change events tell the status of the outgoing process (runnable or not).
- Wake Up events change the status of a process from non runnable to runnable.
- If the event prior to being scheduled out non runnable is a read or a page fault trap, the process is waiting for the associated file. A table must keep track for each process of all the open and mmap operations associating files with file descriptors or memory address ranges.
- If the event prior to being scheduled out non runnable is a waitpid call, the process is waiting for the associated process.
- There are cases where a process is waiting for one of several possible files or processes, with calls to *wait*, *select*, or *poll*. In that case, the waiting time is attributed to the process or file which causes the call to terminate.
- *Poll* and *wait* may also end because of a timeout. The wait time is then attributed to the more general wait for timeout category.

A trace was generated on a freshly rebooted system, with no other activity, while starting up large programs one after another but without any overlap. The waiting time for Galeon [12], for instance, was decomposed into several wait-for-file and wait-for-process components, which accounted for 7.96s of the total 9.27s. The remaining 1.31s is mostly wait-for-timeout (0.89s awaiting a mouse click) and only 0.42s wait-generic. The wait-generic time could be further reduced by identifying in their own category other system call events in the trace, in addition to read, select, poll, waitpid and wait. The wait-generic time is nonetheless available by system call category; .31s out of the .42s total belongs to stat system calls.

```
# Boot with the LTT enabled kernel and start the LTT daemon
[gzip-1.2.4a]$ tracedaemon -ts60 /dev/tracer trace1.trace trace1.proc
TraceDaemon: Output file ready
TraceDaemon: Trace driver open
TraceDaemon: Trace buffers are 1000000 bytes
TraceDaemon: Fetching eip for syscall on depth : 0
TraceDaemon: Daemon wil run for : (60, 0)
TraceDaemon: Done mapping /proc

TraceDaemon: End of tracing

# Analyse the trace output
[gzip-1.2.4a]$ tracevisualizer -a trace1.trace trace1.proc trace1.out

# Look at the trace analysis
[gzip-1.2.4a]$ less trace1.out
...
  Process (1684, 1658): galeon: galeon-bin
...
    User Mode:
      CPU 5.894726
      Elapsed 15.677299
      WaitCPU 2.358053
      WaitFork 0.000002
      BH 811
      KernelTimer 610
      PacketIn 80
      PacketOut 41
      SchedIn 1330
      TimerExpire 136
    Syscall Mode:
      Elapsed/Calls 0.0003861
      CPU 0.308954
      Elapsed 5.119347
      WaitCPU 0.599823
      WaitFile-001.png 0.000453
      WaitFile-002.png 0.000346
      WaitFile-003.png 0.000213
      WaitFile-004.png 0.000216
      WaitFile-005.png 0.000430
...
      sys_fcntl64
        Elapsed/Calls 0.0000054
        CPU 0.000286
        Elapsed 0.000286
        Calls 53

    Traps
      page fault
        Elapsed/Calls 0.0003101
        CPU 0.179146
        Elapsed 2.444663
        WaitCPU 0.448110
        WaitFile-galeon-bin 1.816968
        BH 38
        Calls 7883
        KernelTimer 23
        PacketIn 2
```

```
            PacketOut 1
            PageAlloc 78
            SchedIn 459
            TimerExpire 2

      IRQs
        timer
          Elapsed/Calls 0.0000118
          CPU 0.007725
          Elapsed 0.007725
          Calls 657

        usb-uhci, eth0
          Elapsed/Calls 0.0000263
          CPU 0.003610
          Elapsed 0.003625
          Calls 138
...
Trap entry              1018876582.807382       1684    13      TRAP : page fault;
EIP : 0x08068FC7
Trap exit               1018876582.807387       1684    7
Syscall entry           1018876582.807421       1684    12      SYSCALL : execve;
EIP : 0x080690D7
File system             1018876582.807452       1684    27      EXEC : galeon
Memory                  1018876582.807494       1684    12      PAGE ALLOC ORDER :
0
Memory                  1018876582.807569       1684    12      START PAGE WAIT
Sched change            1018876582.807645       1579    19      IN : 1579; OUT :
1684; STATE : 2
File system             1018876582.807656       1579    20      POLL : 3; MASK : 0
File system             1018876582.807662       1579    20      POLL : 10; MASK :
0
File system             1018876582.807666       1579    20      POLL : 4; MASK : 0
File system             1018876582.807670       1579    20      POLL : 5; MASK : 0
File system             1018876582.807675       1579    20      POLL : 6; MASK : 0
File system             1018876582.807680       1579    20      POLL : 13; MASK :
0
Memory                  1018876582.807688       1579    12      PAGE FREE ORDER :
0
Memory                  1018876582.807699       1579    12      PAGE FREE ORDER :
0
Syscall exit            1018876582.807702       1579    7
Syscall entry           1018876582.807731       1579    12      SYSCALL :
gettimeofday; EIP : 0x0805816D
Syscall exit            1018876582.807733       1579    7
Syscall entry           1018876582.807758       1579    12      SYSCALL : ioctl;
EIP : 0x08056F30
```

The bigger surprise was the 3.5s wait-for-CPU since nothing else was happening on the system. The answer lies in asynchronously cooperating processes. Indeed, Galeon starts a number of helper sub-processes and usually blocks awaiting an answer. This shows up nicely in the wait-for-process category along with the sub-process id. However, Galeon is a complex graphical application which uses the X window server [5] for graphics rendering. The X server is optimised to avoid blocking using asynchronous interactions whenever possible. This allows batching rendering operations, but also parallel execution when the application and the X server are running on separate processors.

### 4.5.2. Identifying Servers Working on behalf of Clients

The last analysis refinement was decomposing the execution time of a server process, labeled CPU, into CPU-for-Client components. A typical client server relationship may be identified by the pattern of a server listening on a socket, a client connecting to the socket, and the server accepting the connection.

The server and client then communicate by reading and writing to the socket until the request is served and the server goes idle. The server then awaits new connections or requests from clients using a select or poll system call.

Assuming that a server, after its initialization phase, is always working on behalf of a client, the server gets into CPU-for-Client $x$ whenever it accepts a connection or reads from a connection to client $x$. This simple heuristic produces interesting results in many cases. However, it does not model cases where requests are grouped. For example, if nine clients make asynchronous requests which are simply buffered by the server, and upon the request from the tenth client, all buffered requests are executed, the time spent serving the last is overestimated.

Complex interactions between clients and servers are even more difficult to automatically recognize in order to attribute serving time appropriately to clients. For instance, Galeon connects to the GNOME configuration daemon, Gconfd, to express interest in a number of configuration parameters, leaving a callback socket address. Gconfd then connects to this callback address and provides the current values for the parameters, and eventually updates to these parameters. Galeon then appears to be serving parameter change requests from client Gconfd.

### *4.6. Discussion*

The Linux Trace Toolkit is one of the best solutions for in depth low level system studies. It is widely used in embedded and real-time systems, and could easily be used in operating systems and real-time programming courses. It could prove very useful to developers and system administrators when strong interactions between applications and the operating system are encountered. It is still pending integration in the main Linux kernel. Thus, a kernel patch and recompilation, or a special kernel, are required on top of the LTT tracing daemon and tracing tools package.

## 5. Conclusion

Traditional tools like Gcov, Gprof, and Strace have existed for years, with many variants in terms of presentation (graphs, colors, even animations) on different platforms. These represent mature, simple and efficient ways to obtain useful information. The appearance of hardware performance counters in microprocessors in the 1990s was the enabling technology for sampling tools like Oprofile providing more precise sampling and several types of samples (cycle count, cache misses...).

Another important enabling technology in the field of system performance analysis is the source code availability of free software. The print statements previously inserted by programmers only in their applications can now be placed everywhere in the system. The Linux Trace Toolkit provides the first 60 interesting tracing points. More importantly, it simplifies the insertion of more tracepoints which will enable a number of sophisticated analysis tools needed to better understand, study and improve complex distributed systems.